\documentclass[superscriptaddress,showpacs,twocolumn]{revtex4}
\usepackage{amssymb}
\usepackage[tbtags]{amsmath}
\usepackage{graphicx}
\usepackage{epsfig,graphicx,times}
\date{\today}

\begin{document}
\title{Engineering quantum pure states of a trapped cold ion beyond the Lamb-Dicke limit}
\author{L.F. Wei}
\affiliation{Frontier Research System,  The Institute of Physical
and Chemical Research (RIKEN), Wako-shi, Saitama, 351-0198, Japan}
\affiliation{Institute of Quantum Optics and Quantum Information,
Department of Physics, Shanghai Jiaotong University, Shanghai
200030, P.R. China }
\author{ Yu-xi Liu}
\affiliation{Frontier Research System,  The Institute of Physical
and Chemical Research (RIKEN), Wako-shi, Saitama, 351-0198, Japan}
\author{Franco Nori}
\affiliation{Frontier Research System,  The Institute of Physical
and Chemical Research (RIKEN), Wako-shi, Saitama, 351-0198, Japan}
\affiliation{Center for Theoretical Physics, Physics Department,
Center for the Study of Complex Systems, The University of
Michigan, Ann Arbor, Michigan 48109-1120}

\begin{abstract}
Based on the conditional quantum dynamics of laser-ion
interactions, we propose an efficient theoretical scheme to
deterministically generate quantum pure states of a single trapped
cold ion without performing the Lamb-Dicke approximation. An
arbitrary quantum state can be created by sequentially using a
series of classical laser pulses with selected frequencies,
initial phases and durations. As special examples, we further show
how to create or approximate several typical macroscopic quantum
states, such as the phase state and the even/odd coherent states.
Unlike previous schemes operating in the Lamb-Dicke regime, the
present one does well for an arbitrary-strength coupling between
the internal and external degrees of freedom of the ion. The
experimental realizability of this approach is also discussed.


\vspace{0.3cm}
\end{abstract}

\maketitle

\section{Introduction}

The engineering of quantum states has attracted considerable
attention in recent years. This in order to test fundamental
quantum concepts, e.g. non-locality, and for implementing various
potential applications, including sensitive detection and quantum
information processing. Recent advances in quantum optics (e.g.,
micromasers, cavity QED) (see, e.g., \cite{Vogel93}) and atomic
physics (ion traps) (see, for instance \cite{Sackett00}) have
allowed a better control of quantum states.

Laser-cooled ions confined in an electromagnetic trap are good
candidates for various quantum-state engineering processes. First,
the trapped ion system possesses relatively long decoherence
times. Second, various interactions including the usual
one-quantum transition Jaynes-Cummings (JC) model and also higher
order non-linear models can be implemented in this system by
simply choosing the applied laser tunings (see,
e.g.,~\cite{Blockley92,Vogel95}). Therefore, a trapped ion driven
by a classical laser field provides the possibility of
conveniently generating various quantum pure states. Indeed,
various engineered quantum states of trapped cold ions have been
studied. The thermal, Fock, coherent, squeezed, and arbitrary
quantum superposition states of motion of a harmonically bound ion
have been investigated \cite{Meekhof96, Leibfried96}. The
manipulation of the entanglement between the external and internal
degrees of freedom of the ion and the realization of a fundamental
quantum logic gate between them has also been demonstrated
experimentally (see, e.g., \cite{Monroe95}).

Most of the previous proposals for engineering the quantum state
of a single trapped cold ion operate in various extreme
experimental conditions, such as the strong Raman excitation or
the weak-coupling Lamb-Dicke (LD) approximations. The former (see,
e.g., \cite{Poyatos96}) requires that the Rabi frequency $\Omega$
characterizing the laser-ion interaction is much larger than the
trap frequency. While the later (see, e.g.,
\cite{Meekhof96,Leibfried96,Matos96}), requires that the coupling
between the external and internal degrees of freedom of the ion is
very weak, i.e., the spatial dimension of the motion of the ground
state of the trapped ion should be much smaller than the effective
wavelength of the applied laser field (see, e.g.,
\cite{Leibfried96}). These approximations can simplify the
laser-ion interaction Hamiltonian to certain solvable models. For
example, in the LD limit the interaction between the internal
states $|s\rangle=\{|g\rangle,\,|e\rangle\}$ and the external
motional harmonic oscillator states $\{|n\rangle;\,n=0,1,2,...\}$
of the ion can be expanded to the lowest order of the LD parameter
$\eta_L$, then the usual JC or anti-JC-type model can be derived.
In addition, the coherent state of the motion of the ion can be
easily generated in those limits. Therefore, an arbitrary quantum
state may be prepared via an atomic interference method by
superimposing a finite number of generated coherent states along a
line. Almost all the quantum-state engineering implementations in
recent ion trap experiments were operated in these limits. Some
meaningful second-order modifications of the these approximations
to the above experimental conditions have been analyzed
theoretically \cite{Zeng99}. However, in general, these limits are
not rigorously satisfied, and higher-order powers of the LD
parameter must be taken into account~\cite{Vogel95}. Indeed, using
the laser-ion interaction outside the LD regime could be helpful
to reduce the noise in the trap and improve the cooling rate (see,
e.g., \cite{Steane98}). Thus, efficiently engineering the quantum
state of the trapped cold ion beyond these limits would be useful.
Arbitrary Fock states can, in principle, be prepared as a dark
motional state of a trapped ion, if the relevant LD parameters can
be set with extreme precision~\cite{Moya00}. More realistically,
reference \cite{Kis01} showed that any pure state, including the
Fock state, can be effectively approximated by the nonlinear
coherent states of the trapped ion. Since these nonlinear coherent
states are one of the motional dark states and are insensitive to
some motional kick effects, the generation of highly excited Fock
states is possible.
Recently, a narrow quadrupole $S_{1/2}$ to $D_{5/2}$ transition at
$729$ nm of a single trapped $^{40}{\rm Ca}^+$ ion has been
successfully manipulated by accurately resolving its vibrational
sidebands \cite{Barton00}. The measured lifetime of the excited
level $D_{5/2}$ is long enough ($\approx 1$\,second) to allow for
a hundred or more quantum operations. Note that the experiments in
~\cite{Barton00} do not strictly operate in the LD regime (with
$\eta\ll 1$), because the corresponding LD parameter is
$\eta\approx 0.25$. Therefore, engineering quantum states of a
single trapped cold ion by exciting various vibrational sidebands
outside the Lamb-Dicke regime is possible to achieve with current
technology.

Based on the exact conditional quantum dynamics for the laser-ion
interaction, including all orders of the LD parameter, in this
paper we propose an efficient scheme for exactly engineering
arbitrary motional and entangled states of a single trapped ion
beyond the LD limit. In this case, all of the target quantum
states are generated deterministically, as any measurement is not
required during the quantum state production or manipulations.

This paper is organized as follows. In Sec. II, we solve exactly
the quantum dynamics for a trapped cold ion driven by a travelling
classical laser beam beyond the LD limit and then introduce some
fundamental unitary operations. By repeatedly using these quantum
operations, in Sec. III we show how to deterministically generate
an arbitrary motional state of a single trapped cold ion. The
preparations of arbitrary entangled states between the external
and internal degrees of freedom are given in Sec. IV. Conclusions
and discussions are given in Sec. V.

\section{Dynamics of a trapped cold ion beyond the Lamb-Dicke limit}

We assume that a single two-level ion is stored in a coaxial
resonator RF-ion trap \cite {Jef}, which provides pseudopotential
oscillation frequencies satisfying the condition $\omega
_{x}\ll\omega _{y,z}$ along the principal axes of the trap. Only
the quantized vibrational motion along the $x$ direction is
considered for the cooled ion \cite {Jef}. The dynamics for such
an ion, driven by a classical travelling-wave light-field of
frequency $\omega _L$ and initial phase $\phi_L$, can be described
by the following Hamiltonian \cite{Blockley92,Cirac96}
\begin{widetext}
\begin{eqnarray}
&&\hat{H}(t)=\hbar \omega \left(\hat{a}^{\dagger
}\hat{a}+\frac{1}{2}\right)+\frac{1}{2} \hbar \omega
_{0}\hat{\sigma}_{z}+\frac{\hbar \Omega }{2}\left\{\hat{\sigma}
_{+}\exp \left[i\eta_L (\hat{a}+\hat{a}^{\dagger })-i(\omega
_Lt+\phi_L)\right]+H.c.\right\}.
\end{eqnarray}
\end{widetext}
The first two terms describe the free motion of the external and
internal degrees of freedom of the ion. Here $\hat{a}^{\dagger }$
and $\hat{a}$ are bosonic creation and annihilation operators of
the atomic vibrational quanta with frequency $\omega$. The Pauli
operators $\hat{\sigma}_{z}$ and $\hat{\sigma}_{\pm }$ are defined
by the internal ground state $|g\rangle$ and excited state
$|e\rangle$ of the ion as $\sigma_{z}=|e\rangle\langle
e|-|g\rangle\langle g|$, $\sigma_{+}=|e\rangle\langle g|$, and
$\sigma_{-}=|g\rangle\langle e|$. These operate on the internal
degrees of freedom of the ion of mass $M$. The final term of
$\hat{H}(t)$ describes the interaction between the ion and the
light field with wave vector $\vec{\kappa}_L$, and initial phase
$\phi_L$. $\Omega$ is the carrier Rabi frequency, which describes
the coupling between the laser and the ion, and is proportional to
the intensity of the applied laser. The frequency $\omega_L$ and
initial phase $\phi_L$ of the applied laser beam are
experimentally controllable~\cite{Cirac96}. Usually, the atomic
transition frequency $\omega _{0}$ between two internal energy
levels is much larger than the trap frequency $\omega$ (e.g., in
the experiments in ~\cite{Barton00}, $\omega _{0}=2\pi\times
4.11\times 10^{11}{\rm kHz}$ and $\omega=2\pi\times 135{\rm
kHz}$). Therefore, for lasers exciting at different vibrational
sidebands with small $k$ values, the LD parameters
\begin{equation}
\eta\,=\,\sqrt{\frac{\hbar\kappa_L^2}{2M\omega}}\,=\,\frac{(\omega_0\pm
k\omega)}{c}\sqrt{\frac{\hbar}{2M\omega}},\,\,\,k=0,1,2,...,
\end{equation}
do not change significantly. Here, the Hamiltonian (1) reduces to
different forms~\cite{Vogel95}
\begin{widetext}
\begin{eqnarray}\label{eq:2}
\hat{H}_{int}=\frac{\hbar\Omega}{2}\times\left\{
\begin{array}{ll}
\vspace{0.3cm}
 \left\{(i\eta)^k
\exp[-\frac{(\eta)^2}{2}-i\phi^r]\,\hat{\sigma}_+
\left(\sum_{j=0}^{\infty}\frac{(i\eta)^{2j}\hat{a}^{\dagger
j}\,\hat{a}^{j+k}}{j!\,(j+k)!}\right)
+H.c. \right\}, &\,\omega_L=\omega_0-k\omega,\\

\vspace{0.3cm} \left\{(i\eta)^k
\exp[-\frac{(\eta)^2}{2}-i\phi^b]\,\hat{\sigma}_+
\left(\sum_{j=0}^{\infty}\frac{(i\eta)^{2j}\hat{a}^{\dagger
(j+k)}\,\hat{a}^j}{j!\,(j+k)!}\right)+H.c. \right\},
&\,\omega_L=\omega_0+k\omega, \\
 \left\{\exp\left(-\frac{\eta^2}{2}-i\phi^c\right)\hat{\sigma}_+
\left(\sum_{j=0}^{\infty}\frac{(i\eta)^{2j}\hat{a}^{\dagger
j}\,\hat{a}^j}{j!\,j!}\right)+H.c. \right\},& \,\omega_L=\omega_0,
\end{array}
\right.\label{eq:10}
\end{eqnarray}
\end{widetext}
in the interaction picture.  The usual rotating-wave approximation
has been made and  all off-resonant transitions have been
neglected by assuming a sufficiently weak applied laser field. The
applied laser beam tuned at the frequency
$\omega_L=\omega_0-k\omega$ ($\omega_L=\omega_0+k\omega$) with
nonzero integer $k$ being denoted as the $k$th red (blue) sideband
line,  because it is red (or blue) detuned from the atomic
frequency $\omega_0$. The line for $k=0$ (i.e.
$\omega_L=\omega_0$) is called the carrier. So, we rewrite the
initial phase $\phi_L$  as  $\phi^{r}$ ($\phi^{b}$, $\phi^{c}$)
for a transition process driven by the red-sideband
(blue-sideband, carrier) laser beam respectively.

Previous discussions (see, e.g., \cite{Matos96}) are usually based
on the LD perturbation approximation to first order in the LD
parameter by assuming $\eta$ to be very small. However, outside
the LD regime the Hamiltonian (\ref{eq:2}) may provide various
quantum transitions between the internal and external states of
the ion. The dynamics for the trapped cold ion governed by the
Hamiltonian (\ref{eq:2}) is exactly solvable \cite{Vogel95,wei02}.
For example, if the external state of the system is initially in
$|m\rangle$ and the internal state is initially in $|e\rangle$ or
$|g\rangle$, then the dynamical evolution of an ion driven by a
red-sideband laser beam with frequency $\omega_L=\omega_0-k\omega$
can be exactly expressed as
\begin{widetext}
\begin{eqnarray}
\left\{\begin{array}{ll} \vspace{0.6cm}
|m\rangle\otimes|g\rangle\,\longrightarrow\,\left\{
\begin{array}{l}
\vspace{0.3cm}
|m\rangle\otimes |g\rangle,\,\,\,\,\,m<k,\\
\cos(\Omega_{m-k,k}t^r_l )|m\rangle\otimes|g\rangle+i^{k-1}
e^{-i\phi^r_l}\sin(\Omega_{m-k,k}t^r_l)|m-k\rangle\otimes|e\rangle,
\,\,\,\,m\geq k, 
\end{array}
\right.\\
|m\rangle\otimes|e\rangle\,\longrightarrow\,\cos(\Omega_{m,k}
t^r_l)|m\rangle\otimes|e\rangle
-(-i)^{k-1}e^{i\phi^r_l}\sin(\Omega_{m,k}t^r_l)|m+k\rangle\otimes|g\rangle,
\end{array}
\right.
\end{eqnarray}
\end{widetext}
with Rabi frequency
$$
\Omega_{m,k}=\frac{\Omega\,
\eta^{k}}{2}\sqrt{\frac{(m+k)!}{(m)!}}e^{-\eta^{2}/2}
\sum_{j=0}^{m}\frac{(i\eta)^{2j}}{(j+k)!}\left(\begin{array}{l} j \\
m \end{array}\right).
$$
Where $m$ is the occupation number of the initial Fock state of
the external vibrational motion of the ion, $t^r_l$ and $\phi^r_l$
are the duration and initial phase of the applied red-sideband
laser beam, respectively. The above dynamical evolution can be
equivalently defined as the $k$th red-sideband ``exciting" quantum
operator
\begin{widetext}
\begin{eqnarray}\label{eq:4}
\hat{R}_{k}^{r}(t^{r}_l)=\left\{
\begin{array}{ll}
|m\rangle|g\rangle\langle m|\langle
g|+\left[\left(1-|\tilde{C}^{r_l}_{m}|^2\right)^{\frac{1}{2}}|m\rangle|e\rangle
+\tilde{C}^{r_l}_{m}\,|m+k\rangle|g\rangle\right]\langle m|\langle e|, \,\,\,\,m<k,\\
\\
\left[\left(1-|C^{r_l}_{m-k}|^2\right)^{\frac{1}{2}}\,|m\rangle|g\rangle+C^{r_l}_{m-k}\,|m-k\rangle|e\rangle\right]
\langle m|\langle g|\\
\hspace{2.0cm}+\left[\left(1-|\tilde{C}^{r_l}_{m}|^2\right)^{\frac{1}{2}}\,|m\rangle|e\rangle
+\tilde{C}^{r_l}_{m}\,|m+k\rangle|g\rangle\right]\langle m|\langle
e|, \,\,\,\,m\geq k,
\end{array}
\right.
\end{eqnarray}
\end{widetext}
with
$$
C^{r_l}_{m}=i^{k-1}e^{-i\phi^r_l}\sin(\Omega_{m,k}t^r_l),\,\,\,\,\,\,
\tilde{C}^{r_l}_{m}=-(C^{r_l}_{m})^*.
$$
The use of $\hat{R}^r_k$ is advantageous because it is compact,
symmetric, and it is simple to iterate. This operator is quite
useful for the generation of quantum states. Analogously, exciting
the motional state of the ion to the $k$th blue-sideband, by
applying a laser of frequency $\omega_L=\omega_0+k\omega$, yields
a unitary blue-sideband ``exciting" quantum operation,
\begin{widetext}
\begin{eqnarray}\label{eq:5}
\hat{R}^{b}_{k}(t^{b}_l)=\left\{
\begin{array}{ll}
\vspace{0.4cm}
\left[\left(1-|C^{b_l}_{m}|^2\right)^{\frac{1}{2}}\,|m\rangle|g\rangle
+C^{b_l}_{m}\,|m+k\rangle|e\rangle\right]
\langle m|\langle g|+|m\rangle|e\rangle\langle m|\langle e|,\,\, m<k,\\
\left[\left(1-|C^{b_l}_{m}|^2\right)^{\frac{1}{2}}\,|m\rangle|g\rangle
+C^{b_l}_{m}\,|m+k\rangle|e\rangle\right]
\langle m|\langle g|\\
\hspace{1.2cm}+\left[\left(1-|\tilde{C}^{b_l}_{m-k}|^2\right)^{\frac{1}{2}}\,|m\rangle|e\rangle
+\tilde{C}^{b_l}_{m-k}\,|m-k\rangle|g\rangle\right] \langle
m|\langle e|,\,\,m\geq k,
\end{array}
\right.
\end{eqnarray}
\end{widetext}
with
$$C^{b_l}_{m}=i^{k-1}e^{-i\phi^b_l}\sin(\Omega_{m,k}t^b_l),\,\,\,\,\,\,\,\,\,
\tilde{C}^{b_l}_{m}=-(C^{b_l}_{m})^*,$$
Here, $t^b_l$ and $\phi_l^b$ are the duration and initial phase of
the applied blue-sideband laser beam, respectively.

Finally, applying a carrier laser pulse with frequency
$\omega_L=\omega_0$, a conditional rotation
\begin{widetext}
\begin{equation}\label{eq:6}
\hat{R}^{c}_{0}(t^c_l)=\left[\left(1-|C^{c_l}_m|^2\right)^{\frac{1}{2}}\,\hat{I}\,+\,C^{c_l}_m\,|e\rangle\langle
g|\,+\,\tilde{C}^{c_l}_m\,|g\rangle\langle e| \right]\otimes
|m\rangle\langle m|
\end{equation}
\end{widetext}
on the internal states of the ion can be implemented. Here,
$\hat{I}=|g\rangle\langle g|+|e\rangle\langle e|$ is the identity
operator, $t^{c}_l$ is the duration of the applied carrier laser
beam, and
$$C^{c_l}_m=-ie^{-i\phi^c_l}\sin(\Omega_{m,0}t^c_l),\,\,\,\,\,\,\,\,\,\,\,
\tilde{C}^{c_l}_{m}=-(C^{c_l}_m)^*,$$  with $$
\Omega_{m,0}=\frac{\Omega}{2}\exp\left(-\frac{\eta^{2}}{2}\right)
\sum_{j=0}^{m}\frac{(i\eta)^{2j}}{j!} \left(\begin{array}{l} j \\
m \end{array}\right).$$

The quantum dynamics of the laser-ion system beyond the LD limit
is conditional. This means that the internal and motional degrees
of freedom are always coupled. The ion states of two degrees of
freedom cannot be operated separately, even if the ion is driven
by the carrier line laser.
Of course, for a given carrier Rabi frequency $\Omega$ (which
depends on the intensity/power of the applied laser beam) and the
LD parameter, the Rabi frequencies $\Omega_{m,k}$ are sensitive to
values of $k$. See, e.g., Fig.~1 for the same laser power but
different LD parameters: $\eta=0.202$~\cite{Meekhof96},
$0.25$~\cite{Barton00}, $0.35,\,0.5,$ and\, $0.9$. As seen in
Figure 1, a smaller $\eta$ corresponds to a larger reduction of
the Rabi frequency for certain $k$ values (e.g., in Fig.~1,
$\Omega_{0,20}/\Omega\sim 10^{-6}$ for $\eta=0.202$). However, for
any fixed value of $k$, larger values of $\eta$ correspond to
larger values of the reduced Rabi frequencies
$2\Omega_{0,k}/\Omega$. Therefore, fast quantum operations can be
obtained outside the LD regime, where $\eta$ is not small.
\begin{figure}\label{fig1}
\vspace{3cm}
\includegraphics[width=12cm]{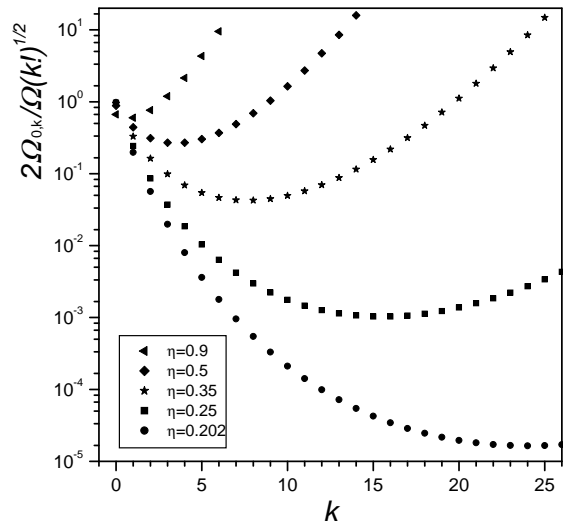}
\vspace{-4cm} \caption{The $k$-dependent Rabi frequency
$\Omega_{0,k}$ for different LD parameters:
$\eta=0.202$~\cite{Meekhof96},
$0.25$~\cite{Barton00},\, $0.35$,\, $0.5$, and \, $0.9$.} 
\end{figure}

In general, any quantum process for the laser-ion system can be
realized by repeatedly applying the above three kinds of
fundamental operations showed in Eqs.~(\ref{eq:4}-\ref{eq:6}).
Which {operation} is applied depends on the laser with the chosen
frequency. Below we will use these fundamental unitary operations
(\ref{eq:4}-\ref{eq:6}) to produce or engineer an arbitrary
quantum state of a single trapped cold ion beyond the LD limit.
The tunable experimental parameters in this process are the
frequency $\omega_L$, wave vector $\vec{\kappa}_L$, and the
duration of the applied laser pulse. The generation of quantum
states described below will start with a common non-entangled
initial state $|\psi_0\rangle=|0\rangle\otimes|g\rangle$; that is,
the trapped ion has been cooled to its motional ground state and
the internal degree of the ion is initially in the low-energy
state $|g\rangle$.

\section{Preparation of an arbitrary motional state of a trapped cold ion}

The first significant quantum state which we want to prepare is
the Fock state of the external vibrational quanta of the ion
\begin{equation}\label{eq:7}
|\psi_1\rangle\,=\,|n\rangle,
\end{equation}
with an arbitrary occupation number $n>0$.  The previous schemes
(e.g. \cite{Meekhof96,Leibfried96,Monroe95, Matos96}) operated in
the LD limit only allowed one-quantum transition process (exciting
and absorbing a single phonon process, respectively). Thus, at
least $(n+1)$ transitions are required between
$|j\rangle\otimes|g\rangle \longleftrightarrow|j\pm
1\rangle\otimes|e\rangle$ to generate the desired state
(\ref{eq:7}). However, for larger values of the LD parameters, the
LD approximation is no longer valid and multi-quantum transitions
must be considered. One can obtain the different phonon
transitions, beyond the LD limit, by choosing an appropriate
driving laser frequency. For example, the quantum transition: $
|0\rangle\otimes|g\rangle \rightarrow|n\rangle\otimes|e\rangle, $
can be realized by choosing a blue-sideband driven laser beam with
frequency $\omega_L=\omega_0+n\omega$. So the desired Fock state
$|n\rangle$ can be easily obtained by using a single blue-sideband
exciting unitary operation $\hat{R}^b_{n}(t^b_0)$ with the
duration $t_{0}^{b}$ satisfying the condition:
$\sin(\Omega_{0,n}t^b_0)=1$. In other words,
$\Omega_{0,n}t^b_0=p\pi/2$ with $p$ an odd integer. Note that the
resulting atomic state evolves to its excited state $|e\rangle$
which may transit to the ground state $|g\rangle$ via spontaneous
emission. In order to avoid the additional excitation of the
desired Fock state due to this emission, an additional operation
$\hat{R}^c_{0}(t^c_n)$ is required to evolve the state $|e\rangle$
into the state $|g\rangle$ keeping the motional  state unchanged,
with the duration $t^c_n$ satisfying the condition $
\sin(\Omega_{n,0}t^c_n)=1. $ Therefore, by sequentially performing
a $\pi/2$ blue-sideband laser pulse and a $\pi/2$ carrier line
laser pulse with initial phases $\phi^b_0$ and $\phi^c_n$,
respectively, a relatively steady target Fock state
$|\psi_1\rangle$ is generated from the vacuum state $|0\rangle$ as
follows
\begin{widetext}
\begin{eqnarray}
|0\rangle\otimes|g\rangle \xrightarrow{\hat{R}^b_{n}(t^b_0)}
 i^{n-1}e^{-i\phi^b_0}
|n\rangle\otimes|e\rangle\, \xrightarrow{\hat{R}^c_{0}(t^c_n)}
-i^ne^{i(\phi^c_n-\phi^b_0)}|n\rangle\otimes|g\rangle.
\end{eqnarray}
\end{widetext}
After these unitary operations, the internal electric state
returns to its initial ground state $|g\rangle$. The target state
$|n\rangle$ can also be prepared by continuously applying a
carrier line operation $\hat{R}^c_{0}(t^c_0)$ and an $n$th
red-sideband exciting operator $\hat{R}^r_{n}(t^r_n)$, i.e.,
\begin{widetext}
\begin{equation}
|0\rangle\otimes|g\rangle
\xrightarrow{\hat{R}^c_{0}(t^c_0)}-ie^{-i\phi^c_0}|0\rangle\otimes|e\rangle
\xrightarrow{\hat{R}^r_{n}(t^r_n)}-(-i)^ne^{i(\phi^r_n-\phi^c_0)}|n\rangle\otimes|g\rangle,
\end{equation}
\end{widetext}
with the durations $t^{c}_{0}$ and $t^{r}_{n}$ satisfying
conditions
$\sin(\Omega_{0,0}t^c_0)=1,\,\,\,\sin(\Omega_{0,n}t^r_n)=1$.
Therefore, two unitary operations are sufficient to
deterministically generate an arbitrary Fock state $|n\rangle$
from the initial ground state $|0\rangle$, if the laser-ion
interaction is operated outside the LD regime by using the chosen
laser beams with desired frequencies.

The more general motional state of the ion which we want to
prepare is the following finite superposition of number states
\begin{equation}
|\psi_2\rangle=\sum_{j=0}^N\,c_j\,|j\rangle,
\hspace{1cm}\sum_{n=0}^N\,|\,c_j\,|^2\,=\,1,
\end{equation}
with $N$ being a finite integer. For a single-mode light field,
this state can be probabilistically generated \cite{Dakna} by
physically truncating the photon coherent state. The efficiency of
generating a quantum state by the quantum truncation may be quite
low due to the necessity of quantum measurements. In the present
quantum-state generation the motional vacuum state $|0\rangle$,
instead of the motional coherent states $|\alpha\rangle$, is given
initially. A quantum-state production scheme, in the LD regime,
for generating the desired state (11) has been proposed in
\cite{Zheng98}. We now extend this scheme to generate the target
state $|\psi_2\rangle$ beyond the LD limit. Indeed, sequentially
using $N+1$ laser pulses with frequencies
$\omega_L=\omega_0,\,\omega_0-\omega,\,\cdots,\,\omega_0-l\omega,\,\cdots,\,\omega_0-N\omega$
and durations $t^c_0,\,t^r_1,\,\cdots,\,t^r_l,\,\cdots,\,t^r_N$,
respectively, the desired state is obtained from the initial state
$|\psi_0\rangle$ by a series of dynamical evolutions showed as
follows:
\begin{widetext}
\begin{eqnarray}\label{eq:11}
|\psi_0\rangle\, &\xrightarrow{\hat{R}^c_0(t^c_0)}&
c_0|0\rangle\otimes|g\rangle-ie^{-i\phi^c_0}\left(1-c_0^2\right)^{\frac{1}{2}}\,|0\rangle\otimes|e\rangle\nonumber\\
&\xrightarrow{\hat{R}^r_1(t^r_1)}&
\left(\sum_{j=0}^1c_j\,|j\rangle\right)\otimes|g\rangle-ie^{-i\phi^c_0}\left(1-\sum_{j=0}^1|c_j|^2\right)^{\frac{1}{2}}|0\rangle\otimes|e\rangle\nonumber\\
 &\cdots&\xrightarrow{\hat{R}^r_l(t^r_l)}
\left(\sum_{j=0}^lc_j|j\rangle\right)\otimes|g\rangle
-ie^{-i\phi^c_0}\left(1-\sum_{j=0}^l|c_j|^2\right)^{\frac{1}{2}}|0\rangle\otimes|e\rangle
\nonumber\\
&\cdots& \xrightarrow{\hat{R}^r_N(t^r_N)}
\left(\sum_{j=0}^Nc_j|j\rangle\right)\otimes|g\rangle=|\psi_2\rangle\otimes|g\rangle.
\end{eqnarray}
\end{widetext}
The duration of the final unitary operations $\hat{R}^r_N(t^r_N)$
has been set to satisfy the condition:
$\sin(\Omega_{0,N}t^r_N)=1$. While, the durations and the initial
phases of other applied laser beams can be used to arbitrarily
prescribe the weights $c_{j}$ of the superposed Fock states
$\{|j\rangle;\,j=0,1,...,N\}$, such as
\begin{widetext}
\begin{eqnarray}\label{eq:12}
c_j=\left\{
\begin{array}{ll}
\vspace{0.3cm}
\left(1-|C_0^{c_0}|^2\right)^{\frac{1}{2}}=\cos(\Omega_{0,0}t^c_0),\,\,\,\,\,\,\,\,\,\,\,\,j=0,\\
\vspace{0.2cm}
C_0^{c_0}\left[\prod_{l=1}^{j-1}(1-|\tilde{C}_0^{r_l}|^2)\right]^{\frac{1}{2}}\tilde{C}_0^{r_j}
=-(-i)^je^{i(\phi^r_j-\phi^c_0)}\sin(\Omega_{0,0}t^c_0)\prod_{l=1}^{j-1}\cos(\Omega_{0,l}t^r_l)
\vspace{-0.0cm}
\sin(\Omega_{0,j}t^r_j),\\
\vspace{0.4cm}
\hspace{10cm}1\leq j\leq N-1,\\
C_0^{c_0}\left[\prod_{l=1}^{j-1}(1-|\tilde{C}_0^{r_l}|^2)\right]^{\frac{1}{2}}
=-(-i)^je^{i(\phi^r_N-\phi^c_0)}\sin(\Omega_{0,0}t^c_0)\prod_{l=1}^{N-1}\cos(\Omega_{0,l}t^r_l),\,\,\,\,j=N.
\end{array}
\right.
\end{eqnarray}
\end{widetext}

Similarly, by sequentially applying a series of blue-sideband
exciting operators $\hat{R}^{b}_j(t^b_j)$ with durations
$t^b_j,\,j=1,2,...,N$, after a carrier line operation
$\hat{R}^c_0(t^c_0)$, we can implement the following deterministic
quantum state generation
\begin{widetext}
\begin{eqnarray}
|\psi_0\rangle\xrightarrow{\hat{R}^b_N(t^b_N)\cdots\hat{R}^b_1(t^b_1)\hat{R}^c_0(t^c_0)}
\left(\sum_{j=0}^Nc'_j|j\rangle\right)\otimes|e\rangle
=|\psi'_2\rangle\otimes|e\rangle,
\end{eqnarray}
\end{widetext}
with
$$
c'_j=\left\{
\begin{array}{ll}
\vspace{0.3cm}
-ie^{-i\phi^c_0}\sin(\Omega_{0,0}t^c_0),\,\,\,\,\,\,\,\,\,\,\,\,\,\,j=0,\\
\vspace{0.1cm} i^{j-1}e^{-i\phi^b_j}\cos(\Omega_{0,0}t^c_0)
\prod_{l=1}^{j-1}\cos(\Omega_{0,l}t^b_l)\sin(\Omega_{0,j}t^b_j),\\
\vspace{0.4cm}
\hspace{5cm}1\leq j\leq N-1,\\
i^{N-1}e^{-i\phi^b_N}\cos(\Omega_{0,0}t^c_0)
\prod_{l=1}^{N-1}\cos(\Omega_{0,l}t^b_l),\,\,j=N.
\end{array}
\right.
$$
Here, the duration $t^b_N$ of the last operation
$\hat{R}^b_N(t^b_N)$ is set as $ \sin(\Omega_{0,N}t^b_N)=1$.

So far, we have shown that the desired superposition of a finite
set of motional Fock states $\{|j\rangle;\,j=0,1,...,N\}$ of the
ion can be generated deterministically from the ground state
$|0\rangle$ by using $N+1$ unitary operations, i.e, a carrier line
operation $\hat{R}^c_0$ and $N$ red-sideband $\hat{R}^r_j$ (or
blue-sideband $\hat{R}^b_j$) exciting operations performed by
using the laser beams with frequencies
$\omega_L=\omega_0-j\omega$\,(or $\omega_L=\omega_0+j\omega$). It
is worth pointing out that the motional state generated above in
Eq.(11) or (13) may be reduced to an arbitrary quantum pure state
of the external vibration of the trapped cold ion, as any weight
$c_j$ in Eq.(12) for the Fock state $|j\rangle$ can be prescribed
arbitrarily. For example, the Pegg-Barnett phase state
\cite{Pegg88}
\begin{equation}
|\theta\rangle_N=\frac{1}{\sqrt{N+1}}\sum_{j=0}^Ne^{ij\theta}\,|j\rangle
\end{equation}
can be obtained from Eqs.~(\ref{eq:11}-\ref{eq:12}) by setting the
initial phases and the durations of the applied laser beams to
satisfy the following conditions
\begin{equation}
\phi^c_0=\frac{\pi}{2},\,\,\,\,\,\,e^{i\phi^r_j}=i^{j-1}e^{ij\theta},
\end{equation}
and
\begin{eqnarray}
c_0&=&\cos(\Omega_{0,0}t^c_0)=\cdots\nonumber\\
&=&\sin(\Omega_{0,0}t^c_0)\prod_{l=1}^{j-1}\cos(\Omega_{0,l}t^r_l)\sin(\Omega_{0,j}t^r_j)=\cdots\nonumber\\
&=&\frac{1}{\sqrt{N+1}},
\end{eqnarray}
which implies that
$t_0^c=2\exp(\eta^2/2)[2n_0\pi+\arccos(1/\sqrt{N+1}\,)]/\Omega$,\,
and
$t_j^r=2\exp(\eta^2/2)[2n_j\pi+\arcsin(1/\sqrt{N-j+1}\,)]/(\Omega\eta^j),\,j\neq
0$,\,with $n_0,n_j=0,1,2,...$.
For example, in the typical experimental system~\cite{Barton00}
where $\omega_0=2\pi\times 4.11\times
10^{11}\,$kHz,\,$\omega=2\pi\times 135$\,kHz,\,$\Omega=50\,$kHz,\,
and $\eta=0.25$, the simplest phase state
$|\theta\rangle_1=(|0\rangle+e^{i\theta}|1\rangle)/\sqrt{2}$ can
be prepared by sequentially applying a resonant laser beam (with
frequency $\omega_L=\omega_0$ and initial  phase $\phi_0^c=\pi/2$)
of the shortest duration $t_0^c\approx 3.24\times 10^{-5}$\,s and
a red-sideband line (with frequency $\omega_L=\omega_0-\omega$ and
initial  phase $\phi_1^r=\theta$) of the shortest duration
$t_1^r\approx 2.6\times 10^{-4}$\,s.

The superposition state generated above may approach some
macroscopic motional quantum states of the ion, if the number $N$
of Fock states $|j\rangle$ is sufficiently large. For example, if
the durations of the applied unitary operations are set to satisfy
the conditions
\begin{eqnarray}
c_0&=&\cos(\Omega_{0,0}t^c_0)=e^{-|\alpha|^2/2}, \,\,\,c_1
=\alpha c_0,\nonumber\\
c_2&=&\alpha^2c_0/\sqrt{2!},\,\,\cdots,\,\,
c_j=\alpha^jc_0/\sqrt{j!}\,\,,\cdots,
\end{eqnarray}
the motional superposition state $\sum_{j=0}^Nc_j|j\rangle$ in the
limit $N\rightarrow\infty$ approaches the usual coherent state
\begin{equation}
|\alpha\rangle=e^{-|\alpha|^2/2}\,\sum_{j=0}^{\infty}\frac{\alpha^j}{\sqrt{j!}}\,|j\rangle.
\end{equation}

Similarly, the usual even or odd coherent states may also be
approached by the superposition motional states generated by
sequentially applying the laser beams with frequencies
$\omega_L=\omega-(2l)\omega,\,l=0,1,2,...$ or
$\omega_L=\omega-(2l+1)\omega$, respectively.

\section{Producing entangled states of a trapped cold ion beyond the Lamb-Dicke limit}

Before, we discussed how to generate a motional quantum state of
the ion. Now, we turn our attention to the problem of how to
control the entanglement between these degrees of freedom. It is
well known that entanglement is one of the most striking aspects
of quantum mechanics and plays an important role in quantum
computation \cite{Monroe95}. A laser-ion system provides an
example for clearly showing how to produce an entanglement between
two different quantum degrees of freedom (see, e.g.,
\cite{Kneer98}). Therefore, the third target state which we want
to prepare is the general entangled state of the internal and
external motion degrees of freedom of the ion
\begin{equation}
|\Psi\rangle\,=\,\sum_{j=0}^{N_g}
d^g_j\,|j\rangle\otimes|g\rangle\,+\, \sum_{j=0}^{N_e}
d^e_j\,|j\rangle\otimes|e\rangle,
\end{equation}
where $\sum_{j=0}^{N_g} d^g_j|j\rangle$ ($\sum_{j=0}^{N_e} d^e_j
|j\rangle$) is the external state associated with the internal
ground (excited) state. Notice that the coefficients $c_j$ in the
state $|\psi_2\rangle$ generated above are prescribed arbitrarily.
Thus, applying an additional conditional operation
$\hat{R}^c_0(t^c_{N+1})$, with duration $t^c_{N+1}$, to the state
$|\psi_2\rangle\otimes|g\rangle$, one may prepare an entangled
state with $N_g=N_e=N$; i.e.,
\begin{eqnarray}
|\psi_2\rangle\otimes|g\rangle\, &\xrightarrow{\hat{R}^c_0}&
\sum_{j=0}^N
\left(d_j^{g}|j\rangle\otimes|g\rangle+d^e_j|j\rangle\otimes|e\rangle\right),
\end{eqnarray}
with
$$
d_j^{g}=c_j\cos(\Omega_{j,0}t^c_{N+1}),\,\,\,\,\,\,\,\,\,\,d_j^{e}=-ic_je^{-i\phi^c_{N+1}}\sin(\Omega_{j,0}t^c_{N+1}).
$$

In the sequence of operations shown above, $N+1$ laser-ion
interactions (a carrier line and $N$ red-sideband/blue-sideband
excitations) are used. We now consider a relatively simple method
to generate the entangled quantum states of the ion. That is, by
alternatively switching the lasers on the first blue-sideband and
the first red-sideband $N$ times, one can generate a typical
entangled state \cite{Kneer98}
\begin{eqnarray}\label{eq:16}
|\Psi'\rangle=\sum_{j=0}^{N_g}d^g_{2j+1}|2j+1\rangle\otimes|g\rangle+
\sum_{j=0}^{N_e}d^e_{2j}|2j\rangle\otimes|e\rangle,
\end{eqnarray}
with the odd (even) -number motional states entangled with the
ground (excited) internal spin states of the ion. Without any loss
of generality, we assume that the ion has been prepared beforehand
in a general non-entangled state
\begin{equation}
|\Psi_0\rangle=\hat{R}^c_0(t^c_0)\,|0\rangle\otimes|g\rangle\,
=d_0^{g_0}\,|0\rangle\otimes|g\rangle
+d_0^{e_0}\,|0\rangle\otimes|e\rangle,
\end{equation}
with
$d_0^{g_0}=\left(1-|C^{c_0}_0|^2\right)^{\frac{1}{2}}$,\,\,\,\,
$d_0^{e_0}=C^{c_0}_0$.

We now first tune the laser beam to the first red-sideband and
thus realize the following operation
\begin{eqnarray}
|\Psi_0\rangle&\xrightarrow{\hat{R}^r_1(t^r_1)}&(d_0^{g_1}\,|0\rangle+d_1^{g_1}\,|1\rangle)\otimes|g\rangle
+d_0^{e_1}|0\rangle\otimes|e\rangle
\nonumber\\
&&=|\Psi_1\rangle.
\end{eqnarray}
Here,
$$
d_0^{g_1}=d_0^{g_0},\,\,\,\,\,\,\,\,
d_1^{g_1}=d^{e_{0}}_0\tilde{C}_0^{r_1},\,\,\,\,\,\,\,\,\,\,
d_0^{e_1}=d_0^{e_0}\left(1-|\tilde{C}_0^{r_1}|^2\right)^{\frac{1}{2}}.
$$
Obviously, the state $|\Psi_1\rangle$ is an entangled state. It
reduces to the Bell-type state
\begin{equation}
|\psi_B\rangle=\frac{1}{\sqrt{2}}(|0\rangle\otimes|e\rangle+|1\rangle\otimes|g\rangle),
\end{equation}
if $\phi_1^r=3\pi/2$ and the durations of the operations
$\hat{R}^c_0(t^c_0)$ and $\hat{R}^r_1(t^c_1)$ are set up properly
such that $d^{e_{0}}_0=1$ and $\tilde{C}_0^{r_1}=1/\sqrt{2}$,
corresponding to the shortest durations $t_0^c\approx 6.48\times
10^{-5}$ s \,and $t_1^r\approx 2.6\times 10^{-4}$ s.

We then tune the laser beam to the first blue-sideband and
implement the evolution
\begin{eqnarray}
|\Psi_1\rangle&\xrightarrow{\hat{R}_1^{b}(t^b_2)}&
\sum_{j=0}^1d^{g_2}_j\,|j\rangle\otimes|g\rangle
+\sum_{j=0}^2d_j^{e_2}\,|j\rangle\otimes|e\rangle\nonumber\\
&&=|\Psi_2\rangle,
\end{eqnarray}

with
$$
d_0^{g_2}=d_0^{g_1}\left(1-|C_0^{b_2}|^2\right)^{\frac{1}{2}},\,\,\,\,\,\,\,\,\,\,
d_1^{g_2}=d_1^{g_1}\left(1-|C_1^{b_2}|^2\right)^{\frac{1}{2}},
$$
and
$$
d_0^{e_2}=d_0^{e_1},\,\,\,\,\,\,\,\,
d_1^{e_2}=d_0^{g_1}C_0^{b_2},\,\,\,\,\,\,\,\,\
d_2^{e_2}=d_1^{g_1}C_1^{b_2}.
$$

Repeating the above-mentioned procedure, we realize the  following
series of evolutions
\begin{widetext}
\begin{eqnarray}
|\Psi_2\rangle\xrightarrow{\hat{R}^r_1(t^r_3)}|\Psi_3\rangle\,\,
\cdots\,\xrightarrow{\hat{R}^r_1(t^b_{2l})}|\Psi_{2l}\rangle
\xrightarrow{\hat{R}^r_1(t^r_{2l+1})}|\Psi_{2l+1}\rangle\,\,
\cdots\,\xrightarrow{\hat{R}^r_1(t^b_{N})}|\psi_{N}\rangle,
\end{eqnarray}
\end{widetext}
with
\begin{eqnarray}
\left\{
\begin{array}{l}
|\Psi_{2l}\rangle
 =\sum_{j=0}^{2l-1}d_j^{g_{2l}}|j\rangle\otimes|g\rangle\,
 +\sum_{j=0}^{2l}d_j^{e_{2l}}|j\rangle\otimes|e\rangle,\nonumber\\
 \\
|\Psi_{2l+1}\rangle=\sum_{j=0}^{2l+1}d_j^{g_{2l+1}}|j\rangle\otimes|g\rangle
+ \sum_{j=0}^{2l}d_j^{e_{2l+1}}|j\rangle\otimes|e\rangle.
\end{array}
\right.
\end{eqnarray}
Here,
\begin{widetext}
\begin{eqnarray}
d_j^{g_{2l}}=\left\{
\begin{array}{l}
\vspace{0.1cm}
d_j^{g_{2l-1}}\left(1-|C_j^{b_{2l}}|^2\right)^{\frac{1}{2}}+d_j^{e_{2l-1}}\tilde{C}_{j+1}^{b_{2l}},\nonumber\\
\vspace{0.4cm}
\hspace{3.0cm}0\leq j\leq 2l-2,\nonumber\\
d_j^{g_{2l-1}}\left(1-|C_j^{b_{2l}}|^2\right)^{\frac{1}{2}},
\,\,\,\,j=2l-1,\,2l,\nonumber
\end{array}
\right.,\,\,d_j^{e_{2l}}=\left\{
\begin{array}{l}
\vspace{0.4cm}
d_j^{e_{2l-1}}, \, \,\,\,\,\,\,\,\,\,\,j=0\nonumber\\
\vspace{0.1cm}
d_j^{e_{2l-1}}\left(1-|\tilde{C}_{j-1}^{b_{2l}}|^2\right)^{\frac{1}{2}}+d_{j-1}^{e_{2l-1}}C_{j-1}^{b_{2l}},\nonumber\\
\vspace{0.4cm}
\hspace{2.8cm}1\leq j\leq 2l-2,\\
d_j^{g_{2l-1}}C_j^{b_{2l}},\,\,\,\,\,\,\,\,\,j=2l-1,\,2l,\nonumber
\end{array}
\right.
\end{eqnarray}
\end{widetext}
and
\begin{widetext}
\begin{eqnarray}
d_j^{g_{2l+1}}=\left\{
\begin{array}{l}
\vspace{0.4cm}
d_j^{g_{2l}},\,\,\,\,\,\,\,\,j=0,\nonumber\\
\vspace{0.1cm}
d_j^{g_{2l}}\left(1-|C_{j-1}^{r_{2l+1}}|^2\right)^{\frac{1}{2}}+d_{j-1}^{e_{2l}}\tilde{C}_{j-11}^{r_{2l+1}},\,  \nonumber\\
\vspace{0.4cm}
\hspace{2.8cm}1\leq j\leq 2l-1,\nonumber\\
d_j^{e_{2l}}\tilde{C}_j^{r_{2l+1}},\,\,\,\,\,\,j=2l,\,2l+1,
\end{array}
\right.,\,\,
 d_j^{e_{2l+1}}=\left\{
\begin{array}{l}
\vspace{0.1cm}
d_j^{e_{2l}}\left(1-|\tilde{C}_{j}^{r_{2l+1}}|^2\right)^{\frac{1}{2}}+d_{j-1}^{g_{2l}}C_{j-1}^{r_{2l+1}},\nonumber\\
\vspace{0.4cm}
\hspace{2.8cm}0\leq j\leq 2l-1,\nonumber\\
d_j^{e_{2l}}\left(1-|\tilde{C}_{j}^{r_{2l+1}}|^2\right)^{\frac{1}{2}},\,\,j=2l,\,2l+1.\nonumber
\end{array}
\right.
\end{eqnarray}
\end{widetext}
Therefore, applying $N$ ($N>0$) pairs of the first red-sideband
and blue-sideband laser beams may generate the desired entangled
state
\begin{equation}
|\psi_N\rangle\,=\,\sum_{j=0}^{N-1}
d_{j}^{g_{N}}\,|j\rangle\otimes|g\rangle\,+\, \sum_{j=0}^{N}
d_{j}^{e_N}\,|j\rangle\otimes|e\rangle.
\end{equation}
If the initial state of the above process of quantum state
manipulation is prepared in $|0\rangle\otimes|e\rangle$ by setting
the duration of the applied carrier line laser to satisfy
condition $|C^{c_0}_0|^2=1$, then the desired entangled state
(\ref{eq:16}), rewritten as
\begin{equation}
|\psi'_N\rangle=\sum_{j=0}^{[\frac{N-1}{2}]}d^{g_N}_{2j+1}|2j+1\rangle\otimes|g\rangle+
\sum_{j=0}^{[\frac{N}{2}]}d^{e_N}_{2j}|2j\rangle\otimes|e\rangle,
\end{equation}
is obtained by the above process. Here $[x]$ is the
largest integer less than $x$.

\section{Discussions and Conclusions}

Based on the conditional quantum dynamics for laser-ion
interactions beyond the Lamb-Dicke limit, we have introduced three
kinds of unitary operations: the simple rotations of the internal
states of the ion, the arbitrary red-sideband, and blue-sideband
exciting operations. These unitary operations can be performed
separately by applying the chosen laser beams with the relevant
tunings. In general, any quantum state  of the trapped cold ion
can be generated deterministically by making use of these unitary
operations selectively.
Like some of the other schemes presented previously, several laser
beams with different frequencies are also required in the present
scheme.
Tunable lasers provide the tool to create several types of quantum
states of trapped ions.

Compared with previous approaches (see, e.g., \cite{Monroe95,
Matos96}) operated in the LD regime, the most important advantage
of the present scheme is that the operations are relatively
simple, since various laser-ion interactions may be easily used by
choosing the tunings of the applied laser beams. For example, at
least $n$ operations are required in the previous schemes to
generate the state $|n\rangle\otimes|g\rangle$ from the initial
state $|0\rangle\otimes|g\rangle$, as the dynamical process of the
multiquantum motional excitation is negligible in the LD regime.
However, we have shown here that only two unitary operations
beyond the LD limit are sufficient to engineer the same quantum
state. In addition, the generated superposition motional states
and the entangled states of the ion are universal and thus may
reduce to the various desired special quantum states. The reason
is that the weights of the superposed Fock states can be adjusted
independently by controlling the relevant experimental parameters;
e.g., the durations, initial phases and frequencies of the applied
laser beams. Several typical macroscopic quantum states of the
motion of the ion, e.g., the Pegg-Barnett phase state, the
coherent state, and the even and odd coherent states, etc. can be
created or well-approximated, if the number of the superposition
Fock states is sufficient large.

We now give a brief discussion for the realizability of our
approach.

First, the present quantum manipulations need to resolve the
vibrational sidebands of the ion trap. In fact, it is not
difficult to generate the desired laser pulse with sufficiently
narrow line-width with current experimental technology. For
example, the line-width ($1$ kHz) of the laser beams (at $729$ nm)
used in Ref.~\cite{Barton00} to drive the trapped ion
$^{40}$Ca$^+$ is very small, corresponding to a resolution of
better than $\Delta\nu/\nu=2.5\times 10^{-12}$. This line-width is
also much smaller than the frequency of the applied trap ($138$
kHz). Thus, the vibrational sidebands can be well resolved.
The expected initial  phases of the applied laser beams can be
controlled by switching different signal paths~\cite{Gulde03}.
During the very short durations (e.g., $\lesssim 10^{-4}$ s) for
implementing the expected quantum operations, the applied laser
beams (generated by the Ti: sapphire laser) are sufficient stabile
(e.g., the corresponding initial phase-diffusion and
frequency-drift times may reach to, e.g., $10$ ms~\cite{Chou94}
and bandwidth $\lesssim 1kHz$ in $1$ s averaging
time~\cite{Roos99}, respectively.). In fact, the proposed
engineering scheme could also, in principle, be used for Raman
excitation, where the phases of the applied laser beams can be
well controlled (see,
e.g.,~\cite{Sackett00,Meekhof96,Leibfried96}).

Second, in the present scheme, an arbitrary Fock state can be
prepared by using only two operations (see, e.g., Eqs. (9) and
(10)). The duration of operation depends on the value of $k$, once
the LD parameter and the intensity of the applied laser beam are
given. The Rabi frequency does not significantly reduce for large
LD parameters (e.g., $\eta\gtrsim 0.5$). However, for small values
of $\eta$ (e.g., $\eta\lesssim 0.25$), the Rabi frequency
decreases fast for sufficiently small values of $k$. Small values
of the Rabi frequency correspond to a long duration of quantum
operations, and the allowed number of operations will be reduced.
For example, if $\eta=0.25,\,\Omega=2\pi\times 50$ kHz, then the
duration of the transition $|g\rangle\longrightarrow|e\rangle$ is
$t^c_0\approx 10\,\mu$s. Adding a $\pi/2$ pulse with duration
$t^r_1\approx 40\,\mu$s (or $t^r_{10}\approx 0.3$\,s), the Fock
state: $|1\rangle$ (or $|10\rangle$) can be generated. Note that,
compared to the $t^r_1,\,t^c_0$, the durations of operations
$\hat{R}^r_{10}$ is relatively long, as the Rabi frequencies are
relatively small; $\Omega_{0,10}/\Omega\approx 2\times 10^{-5}$
for the same laser intensity (see, Fig.~1). In principle, these
decreased Rabi frequencies can be effectively compensated by
enhancing the powers (i.e., intensities) of the applied laser
beams. In fact, the power of the laser applied to drive the
trapped cold ion is generally controllable (e.g., the Ti: sapphire
laser used in experiment~\cite{Barton00} is adjustable in the
range from a few $\mu$W to a few hundred mW). Therefore, the
corresponding durations can be shorter by $2\sim 5$ orders of
magnitude. For example, for $\eta=0.25$, if the power of the
applied laser beam is adjusted from a few $\mu$W to a few mW, the
Rabi frequency $\Omega_{0,10}$ of the transition
$|0\rangle\otimes|e\rangle\rightarrow|10\rangle|g\rangle$ can be
enhanced to the same order of magnitude of the carrier Rabi
frequency $\Omega\,(=2\pi\times 50)$\,kHz. The duration of the
corresponding quantum operation is then shortened to $10^{-5}$s.
The smaller LD parameters $\eta$ correspond to lasers with larger
adjustable power ranges; e.g., for $\eta=0.202$, the adjustable
power range should be five orders of magnitude larger. Therefore,
it is difficult to realize transitions with higher $k$ in the LD
regime, where $\eta\ll 1$.

Finally, the generation of the macroscopic superposed Fock states
is limited in practice by the existing decays of the vibrational
and atomic states. The lifetime of the atomic excited state
$|e\rangle$ reaches up to $~1$ s \cite{Barton00} allows, in
principle, to perform $10^3\sim 10^4$ manipulations. Also, the
recent experiment~\cite{Roos99} showed that coherence for the
superposition of $|n=0\rangle$ and $|n=1\rangle$ was maintained
for up to $1$ ms. Usually, the lifetime (i.e., relaxation time) of
the state $|1\rangle$ should be longer than this dephasing time.
Therefore, roughly say, preparing a superposition (e.g., phase
state) from ground state $|n=0\rangle$ to the excited motional
Fock states $|n\rangle$ with $n>10$ is experimentally possible, as
the durations of quantum operation are sufficiently short, e.g.,
$<10^{-4}$s. Improvements might be expected by considering the
more realistic dynamics \cite{Kis01} that includes the decays of
the excited atomic and motional states.

\section*{Acknowledgments}
This work was supported in part by the National Security Agency
(NSA) and Advanced Research and Development Activity (ARDA) under
Air Force Office of Research (AFOSR) contract number
F49620-02-1-0334, and by the National Science Foundation grant No.
EIA-0130383.

\end{document}